\documentclass[twocolumn]{aastex63}

\newcommand\lsim{\mathrel{\rlap{\lower4pt\hbox{\hskip1pt$\sim$}}
\raise1pt\hbox{$<$}}}
\newcommand\gsim{\mathrel{\rlap{\lower4pt\hbox{\hskip1pt$\sim$}}
\raise1pt\hbox{$>$}}}

\newcommand{\bc}{\mathrm{bc}}
\newcommand{\msun}{M_\odot}

\newcommand{\svbc}{\sigma_{v_{\rm bc}}}
\newcommand{\vbc}{v_{\rm bc}}

\received{X}
\revised{Y}
\accepted{Z}

\submitjournal{ApJ}

\shorttitle{SIGO cooling}
\shortauthors{Chiou et al.}

\begin{document}

\title{The Supersonic Project: To cool or not to cool Supersonically Induced Gas Objects (SIGOs)?}

\correspondingauthor{Yeou S. Chiou}
\email{yschiou@physics.ucla.edu}

\author[0000-0003-4962-5768]{Yeou S. Chiou}
\affil{Department of Physics and Astronomy, University of California, Los Angeles, CA 90095\\}
\affil{Mani L. Bhaumik Institute for Theoretical Physics, Department of Physics and Astronomy, UCLA, Los Angeles, CA 90095, USA\\}

\author[0000-0002-9802-9279]{Smadar Naoz}
\affil{Department of Physics and Astronomy, University of California, Los Angeles, CA 90095\\}
\affil{Mani L. Bhaumik Institute for Theoretical Physics, Department of Physics and Astronomy, UCLA, Los Angeles, CA 90095, USA\\}

\author[0000-0001-5817-5944]{Blakesley Burkhart}
\affiliation{Department of Physics and Astronomy, Rutgers, The State University of New Jersey, 136 Frelinghuysen Rd, Piscataway, NJ 08854, USA \\}
\affiliation{Center for Computational Astrophysics, Flatiron Institute, 162 Fifth Avenue, New York, NY 10010, USA \\}

\author[0000-0003-3816-7028]{Federico Marinacci}
\affiliation{Department of Physics \& Astronomy, University of Bologna, via Gobetti 93/2, 40129 Bologna, Italy\\}
\affiliation{Harvard-Smithsonian Center for Astrophysics, 60 Garden Street, Cambridge, MA 02138, USA\\}

\author[0000-0001-8593-7692]{Mark Vogelsberger}
\affil{Department of Physics and Kavli Institute for Astrophysics and Space Research, Massachusetts Institute of Technology, Cambridge, MA 02139, USA\\}

\begin{abstract}
Supersonically Induced Gas Objects (SIGOs) primarily form in the early Universe, outside of dark matter halos due to the presence of a relative stream velocity between baryons and dark matter. These structures may be the progenitors of globular clusters. Since SIGOs are made out of pristine gas, we investigate the effect of atomic cooling on their properties. We run a suite of simulations by using the moving-mesh code {\sc arepo}, with and without baryon-dark matter relative velocity and with and without the effects of atomic cooling. We show that SIGO's density, temperature, and prolateness are determined by gravitational interactions rather than cooling. The cold gas fraction in SIGOs is much higher than that of dark matter halos. Specifically, we show that the SIGO's characteristic low temperature and extreme high gas density forges a nurturing ground for the earliest star formation sites. 
\end{abstract}


\section{Introduction}\label{sec:Intro}

One of the fundamental tenets of the Lambda cold dark matter ($\Lambda$CDM) cosmology is the hierarchical buildup of structure in a bottom-up fashion \citep[e.g.,][]{rees77,silk77,white78,blumenthal84}. Smaller structures collide and build bigger structures. After inflation, dark matter (DM) in the Universe started this buildup by clustering gravitationally. However, since baryonic matter and radiation were tightly coupled, it was not until after the time of recombination that baryonic overdensities could grow. After decoupling, the baryons felt the gravitational potential wells of DM overdensities, as well as their own, and subsequently their overdensities grew. 

 \citet{TH}  noted that the evolution during these early times yielded a relative velocity between the DM and baryons, the so called ``stream velocity.'' Although formally a second order effect, the stream velocity has far-reaching consequences on a wide variety of cosmological phenomena \citep[e.g.,][]{Stacy+10,miao11,greif11,fialkov2012,naoz11,naoz12,oleary12,richardson13,tanaka14}. For example, this relative velocity may have nontrivial effects on the cosmological 21-cm signal \citep[e.g.][]{dalal10,Visbal+12,McQuinn+12, fialkov+14, munoz19}, the formation of primordial black holes \citep[e.g.,][]{tanaka13,tanaka14,latif14,hirano17,schauer17}, and even for primordial magnetic fields \citep{naozyoshida13}. 

In particular, it was recently suggested that the stream velocity can result in gas-dominated structures \citep{naoznarayan14}, and that these Supersonically Induced Gas Objects (SIGOs) could be the progenitors of globular clusters \citep{naoznarayan14,popa,chiou18, chiou+19}. 
SIGOs arise naturally as a consequence of the stream velocity effect \citep{TH}. In particular, this relative velocity introduces a phase shift between DM and baryons, which can lead to baryon overdensity peaks collapsing outside the virial radius of their parent DM halo \citep{naoznarayan14}. 

We previously suggested that SIGOs may be a viable nurturing ground for early star formation \citep{chiou+19}. Specifically, we used a combination of simulation and semi-analytical modeling to follow the SIGOs from birth to present day. We showed that present day SIGOs share characteristics with observed globular clusters.  While in \citet{chiou+19} we studied star formation and the connection between SIGOs and globular clusters, here we focus specifically on the effect that atomic cooling has on SIGOs and run detailed comparisons between simulations with cooling turned on versus turned off.

Cooling plays a vital role in the formation of the first stars in the Universe, the so-called  Population  III  (Pop  III)  stars  (see \cite{wise+12,wise12}, \cite{Bromm13} and references within for a review). Pop III stars are formed in a metal-free environment and are typically assumed to exist within a DM halo of $\sim 10^6$~M$_\odot$. These DM halos contain gas whose chemical composition is a primordial mix of hydrogen and helium, after they reached their cosmological Jeans mass \citep{barkanaloeb2001}, or filtering mass \citep[the highest mass at which the gas pressure still manages to balance gravity, e.g.,][]{Naoz+07,Naoz+09,naoz11}. To model the formation of stars, a sub-grid procedure is usually adopted and a density threshold that expresses the  balance  between  gravity  and  thermal  pressure is considered \citep[e.g.,][]{Bromm+99,Bromm+02,Yoshida+06,Yoshida+07,Wollenberg+20}.
These stars formed in pristine gaseous clouds that were heated and shocked as they fell into DM minihalos ($10^5 - 10^7 \msun$). These objects would become pressure supported unless cooling via atomic and molecular hydrogen allowed them to collapse to such densities that stars could form.

In the context of cosmological simulations, cooling has been taken into account in a variety of ways with different cooling channels \citep[e.g.,][]{Abel+02,Bromm+02,Reed+05,Yoshida+06,Stacy+10,Glover13,Vogelsberger+13, Xu+16,Sarmento+18}. The standard assumptions are that hydrogen and helium are in photo-ionisation equilibrium with a uniform, but time-varying, UV background and that metals are in collisional ionisation equilibrium.

The first stage of cooling for pristine non-molecular gas is atomic hydrogen cooling. This cooling channel is typically more effective for gas with temperatures $>10^4$~K \citep[e.g.,][]{barkanaloeb2001}. Once the gas becomes metal-enriched it can further cool via molecular cooling \citep[e.g.,][]{Saslaw+67,McGreer+08}. In this case $H_2$ can form on dust grains, (or via a three-body $H_2$ formation process).  $H_2$ cooling, thus, becomes effective for a gas cloud when the number density reaches above $10^4$ cm$^{-3}$ and further allows cooling down to temperatures of $300$~K \citep{Abel+98,Abel+02,Anninos+97,wise12}. Since we focus on the very first gas structures, which include pristine gas, we target the atomic cooling channel.

Recently, \citet{schauer+19} incorporated the stream velocity effect between DM and baryons in the early Universe with an advanced chemistry network. Focusing on classical DM halos, they showed the minimum halo mass that allows star formation is significantly increased and that the cooling of gas is suppressed. They, however, did not investigate the effect of cooling on SIGOs properties. In fact, no study until now has  done that. Although SIGOs average temperature is below $10^4$~K, they are coupled to the gravitational potential of their surroundings and live in a complex, turbulent medium. Therefore, cooling the gas may affect their physical properties (such as their morphology and overdensities), since it influences the overall gas dynamics. Thus, here we study, for the first time, the potential effects that cooling may have on the SIGOs’ physical properties. 
 
In this paper, we investigate the effect of cooling on SIGO morphology and potential star forming properties by analyzing the cosmological simulations in \citet{chiou+19} with cooling and UVB shielding taken into account.
This paper is organized as follows: in Section \ref{sec:Sim} we detail the simulations. In Section \ref{sec:cooling} we describe how cooling affects the characteristics and morphological properties of SIGOs and we wrap up with a discussion and our conclusions in Section \ref{sec:conclusions}.

Throughout this paper, we assume a $\Lambda$CDM cosmology with $\Omega_{\Lambda} = 0.73$, $\Omega_m= 0.27$, $\Omega_B= 0.044$, $\sigma_8= 1.7$, and $h = 0.71$. All the quantities that we analyze in this paper are expressed in physical units.

\section{Simulations}\label{sec:Sim}
\subsection{Cooling and non-cooling runs}

We perform a suite of cosmological simulations using the moving-mesh code {\sc arepo} \citep{springel10}. {\sc arepo} has the advantage that the moving mesh can continuously and automatically adapt to mass clustering. We run a  $2$~cMpc (co-moving Mpc) box with $512^3$~DM particles and $512^3$ Voronoi mesh cells. At this resolution, the DM (Voronoi mesh cell) mass is $1.9\times10^3$~M$_\odot$ ($360$~M$_\odot$). We employ a modified version of the {\sc cmbfast} \citep{seljak96} code to generate the transfer functions used for the initial conditions. These transfer functions take into account the first-order correction of the scale-dependent temperature fluctuations  \citep[as shown in][]{NB}  and also include the stream  velocity  effects  following  \citet{TH} and \citet{tseliakhovich11}. We begin our simulation at $z=200$ and present results for $z=20$ \citep[which allows us to overcome numerical problems at smaller redshift for this small box, as explained in details in][]{popa,chiou18,chiou+19}.  We note that our simulations naturally include different transfer functions for the DM and baryon components, which play an important role in determining the gas fraction and overdensity growth rate \citep[e.g.,][]{Naoz+07,naoz11,naozyoshida13, park+20}.

 We include radiative cooling, with and without stream velocity effects. However, we do not include explicit star formation or feedback\footnote{ Note that in \citet{chiou+19},  star formation was treated with a semi-analytical approach.}. {\sc arepo}'s basic cooling module is composed by a primordial chemistry and cooling network  of the evolution of species H, H$^+$, He, He$^+$, He$^{++}$ and e$^-$ in equilibrium with a photoionizing background. The background is assumed to be spatially constant, but redshift dependent.
The gas cooling and heating rates are calculated as a function of gas density, temperature, and redshift\footnote{Note that while all cooling rates include self-shielding corrections, these  do not apply above redshift of $6$, and thus do not contribute for the cooling of the $z=20$ objects.}  \citep[See][and references therein for further details on the numerical implementation of these processes]{Vogelsberger+13}. 

 As a control, we also ran simulations without cooling, and with and without stream velocity effects \footnote{We note that the simulations in \citet{chiou+19} are denoted here as V0Cool and V2Cool}.
 We perform a simulation with a stream velocity of $2~\svbc$ and one without stream velocity. The stream velocity effect is implemented as a uniform boost in the x-direction for the baryons. 
 We note that, while the $2\svbc$ with cooling simulation is presented in \citet{chiou+19}, here we focus on the comparison between cooling and not cooling and with and without the stream velocity.  

Overall we present here a total of $3$ simulation results at $z=20$, with and without cooling, and with and without stream velocity, with the aforementioned parameters\footnote{We omit the results of the no cooling and no stream velocity results (v0Uncool) to avoid clutter. These cases are studied extensively in the literature in the context of small boxes, \citep[e.g.,][]{Naoz+09,naoz11,popa,chiou18}.}.
A summary of the simulations performed is given in Table \ref{table:runs}. The naming convention is as follows. 0 or 2 represents $\vbc = 0, 2 ~\svbc$ and `Cool' or `Uncool' denote whether cooling was turned on or switched off.

\begin{table}
\begin{tabular}{|l | c  |  c | }
\hline 
Name &  $\vbc (\svbc)$	   & Cooling  \\

\hline
v0Cool    &  $0$	  & Yes   \\

v2Cool &   $2$	 &  Yes  \\

v2Uncool &    $2$	 &    No   \\

\hline
   \end{tabular}
           \caption{Details of the simulations performed. For all simulations we have $2$~cMpc box, $512^3$ DM particles, and $512^3$ gas cells. For this set of parameters we have that the mass of the DM particle is $m_{\rm DM}=1.9 \times 10^3 \msun$ whereas the mass of a gas cell is $m_{\rm gas}=360 \msun$. }\label{table:runs}
\vspace{-0.2cm}
\end{table}

 \begin{figure*}
    \includegraphics[width=.32\textwidth]{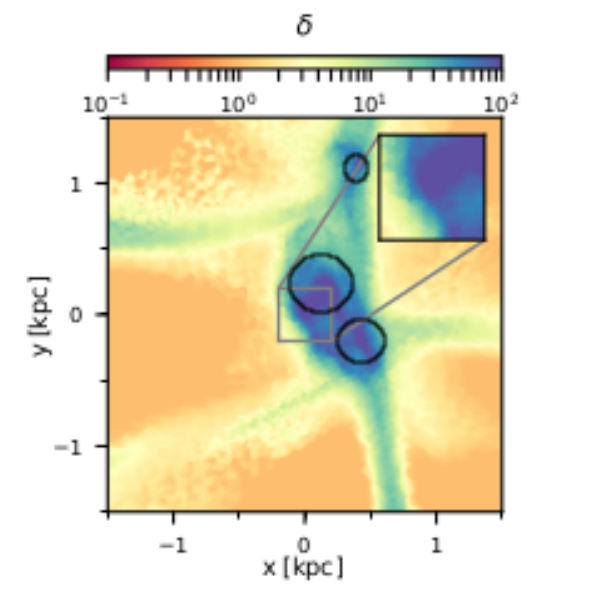}
    \includegraphics[width=.32\textwidth]{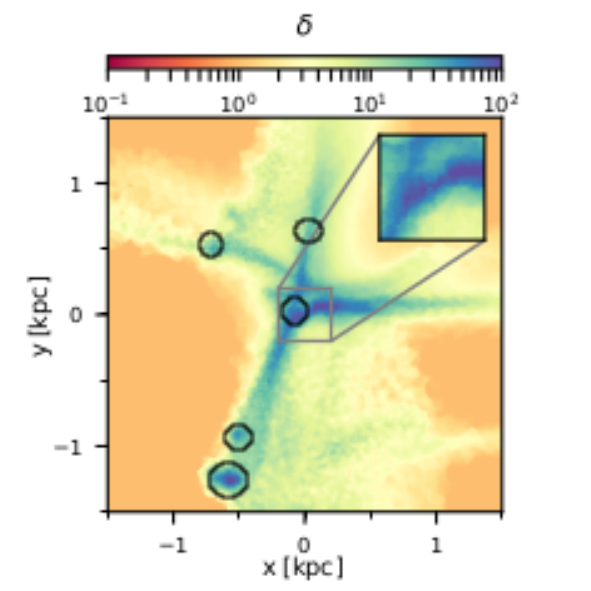}
    \includegraphics[width=.32\textwidth]{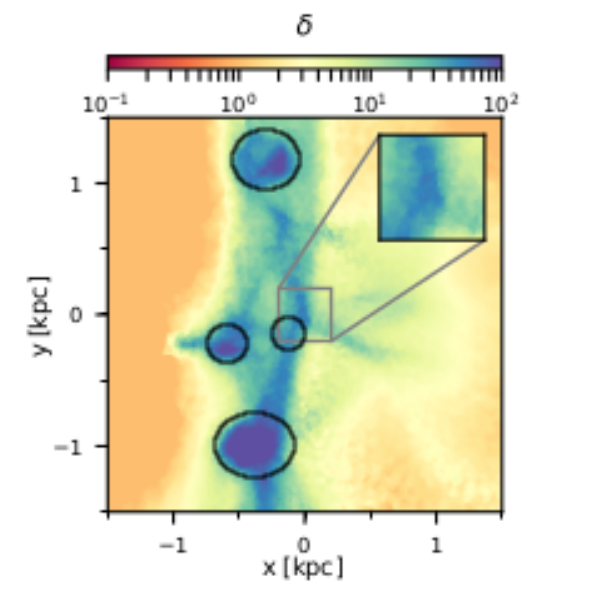}
    \caption{Projected overdensities of various SIGO environments in v2Cool. SIGOs are located in the center of each panel with a zoom-in to the center region showed in each subpanel.  Notice that in regions at which the SIGOs are nested, large by a factor of a few, have projected $\delta>1$, already at $z=20$.
    DM/G are identified by the black circles. In the left and right panels we detect one SIGO, while  the middle panel depicts a ``binary'' SIGO configuration. Note that in projection the SIGOs seem to overlap the classical DM/G. However, as shown in Figure \ref{fig:SIGO_3d}, these SIGOs are detached from the DM/G.}
    \label{fig:over}
\end{figure*}
\begin{figure}
    \centering
    \includegraphics[width=0.5\textwidth]{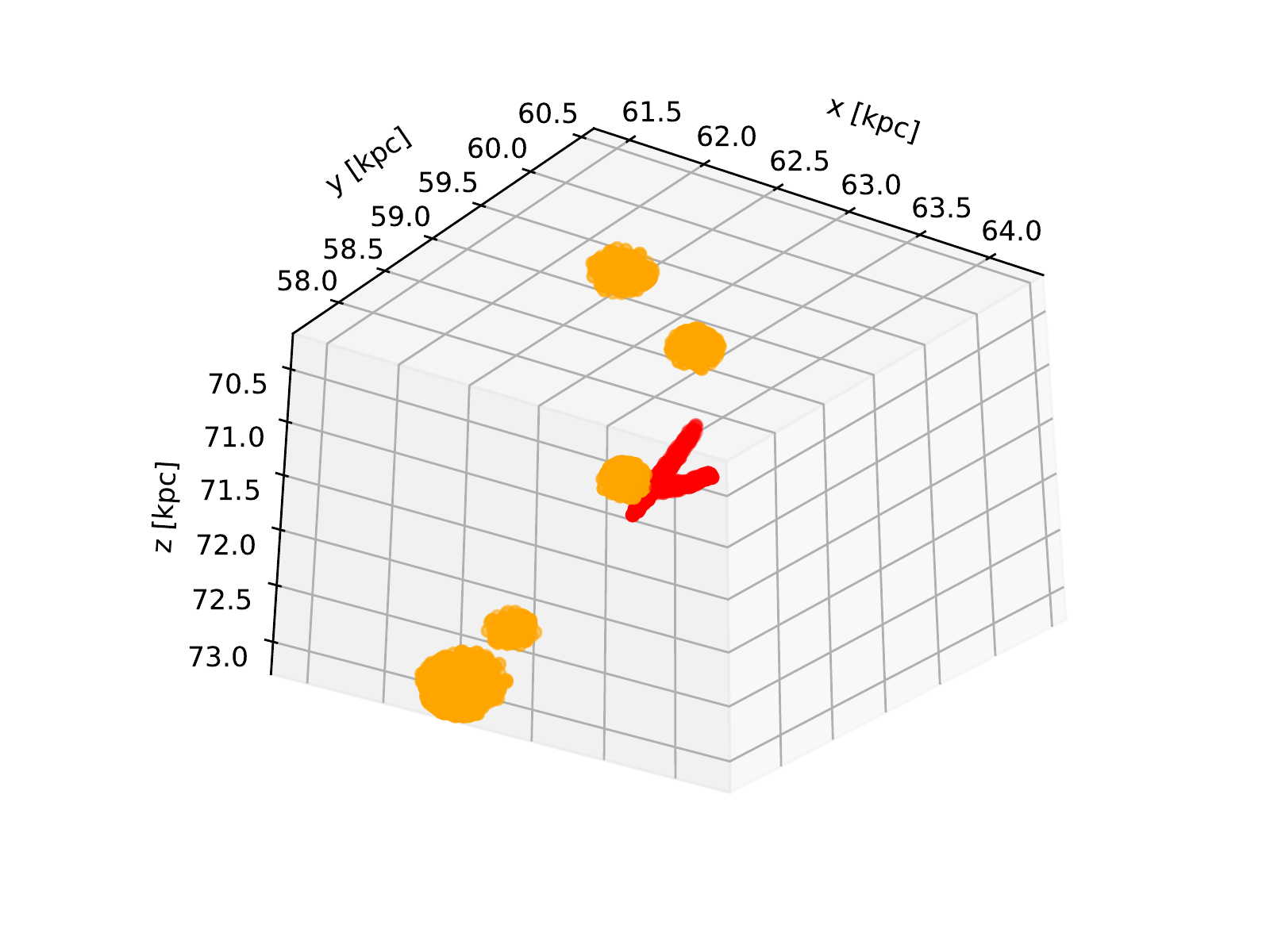}
    \caption{SIGOs corresponding to the middle panel in Figure \ref{fig:over}. There are two SIGOs in red and the surrounding DM/G are in orange. Note the deviation form sphericity as shown in \citet{popa} and \citet{chiou18}. To generated this Figure we ignore ambient particles and only consider DM and gas particles that are bound to an object.} 
    \label{fig:SIGO_3d}
\end{figure}

\subsection{Object classification}

Following \citet{popa} and \citet{chiou18, chiou+19}, we define the following objects in our simulation: DM-Primary/Gas-Secondary (DM/G) and Supersonically Induced Gas Objects (SIGO). These objects come from a Friends-of-Friends (FOF) halo finding algorithm with a linking length of 0.2 times the mean particle separation. We define the DM/Gs by running the FOF algorithm first on the DM, and then adding gas cells in a secondary linking step \citep{dolag09}. We only consider the DM/G to be all particles within a sphere of radius of $R_{200}$ of the center of mass of the FOF group with the further constraints that they contain over 300 DM particles and over 100 gas cells. SIGOs are gas-only FOF groups whose center of mass is outside the virial radius of the closest DM/G and have gas fractions $>40\%$. SIGOs must contain more than 32 gas cells and are fitted to tight ellipsoids. Notice that by definition, SIGOs only appear in runs with nonzero stream velocity.

\begin{figure*}
    \centering
    \includegraphics[width=.32\textwidth]{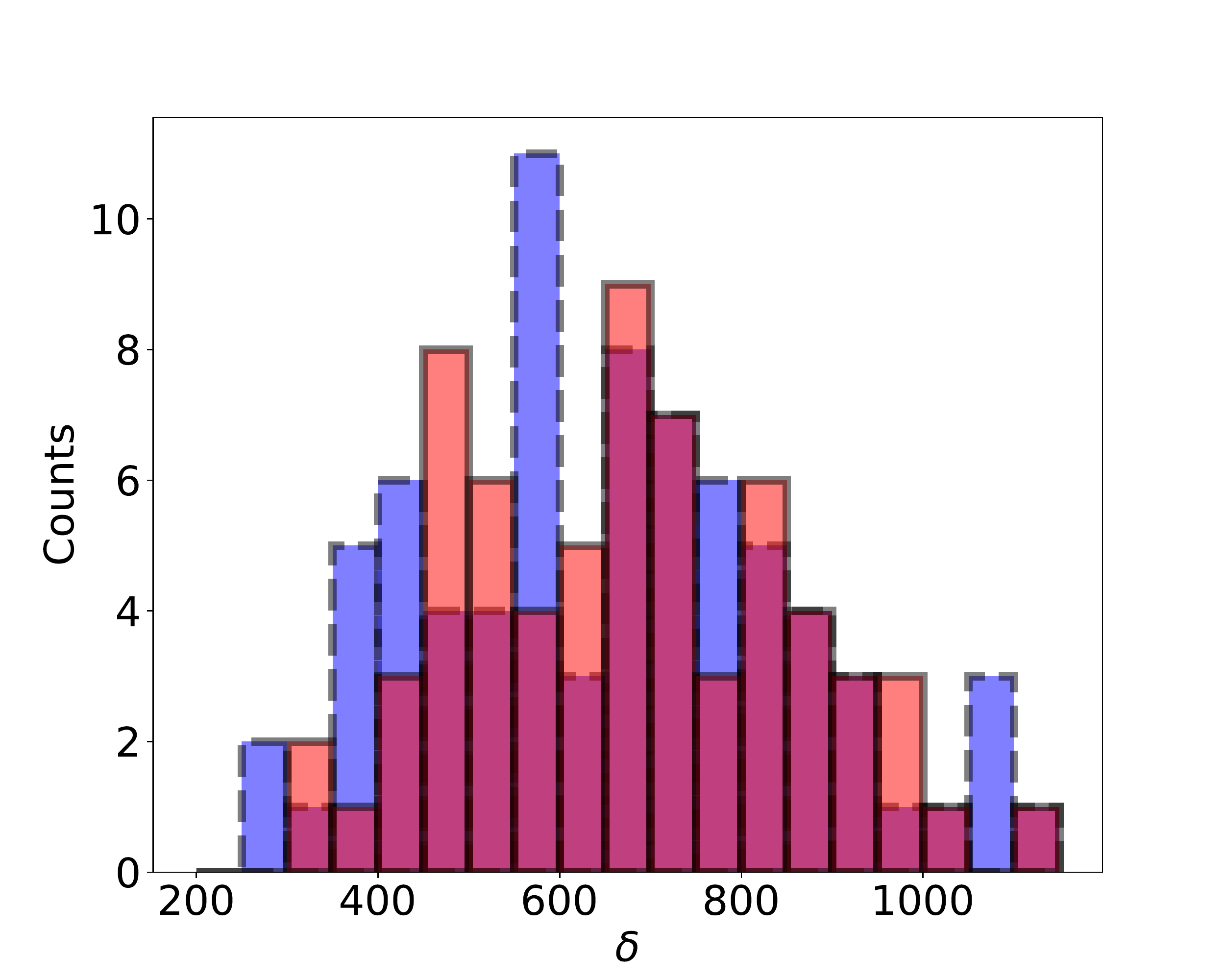}
    \includegraphics[width=.32\textwidth]{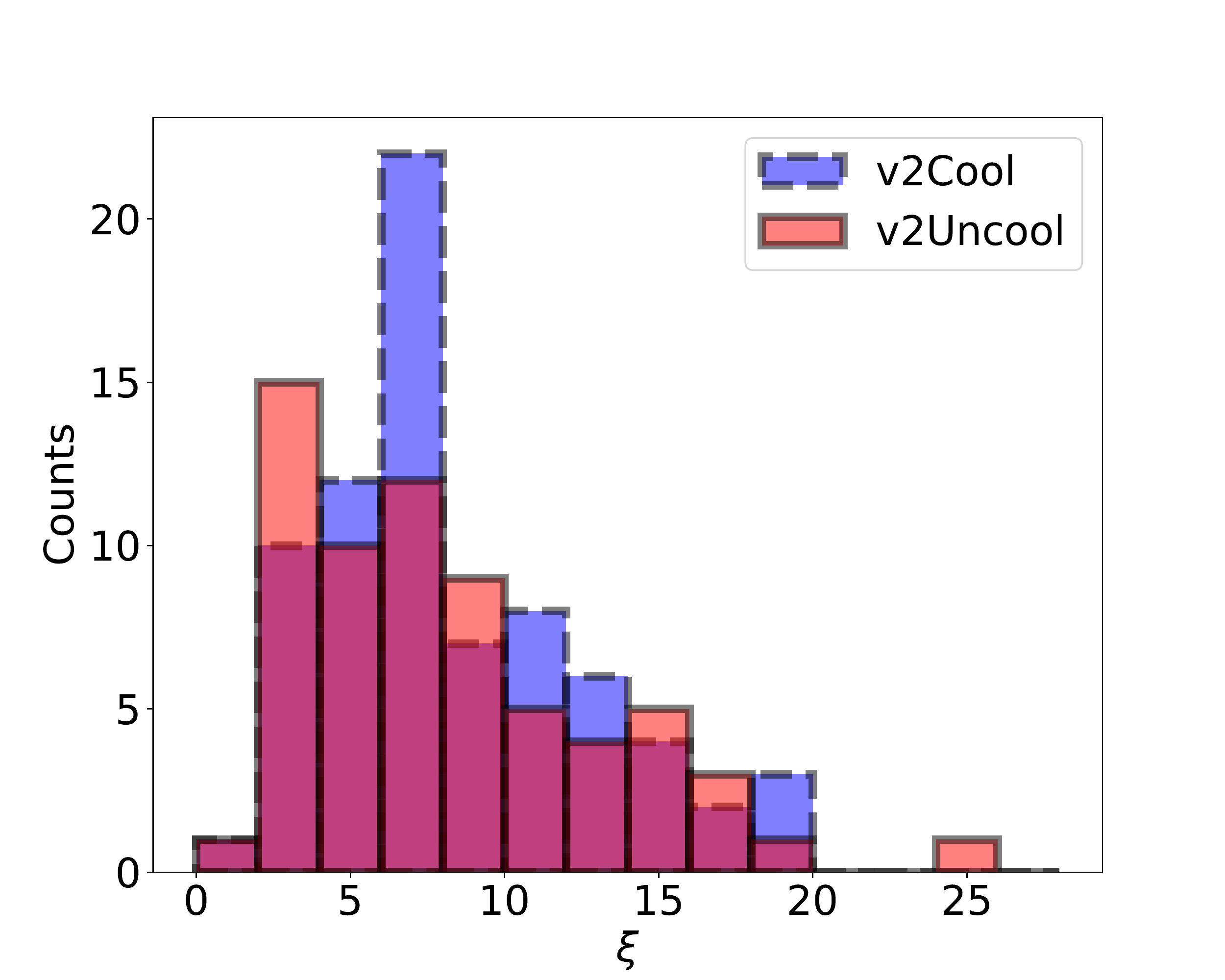}
    
    \includegraphics[width=.32\textwidth]{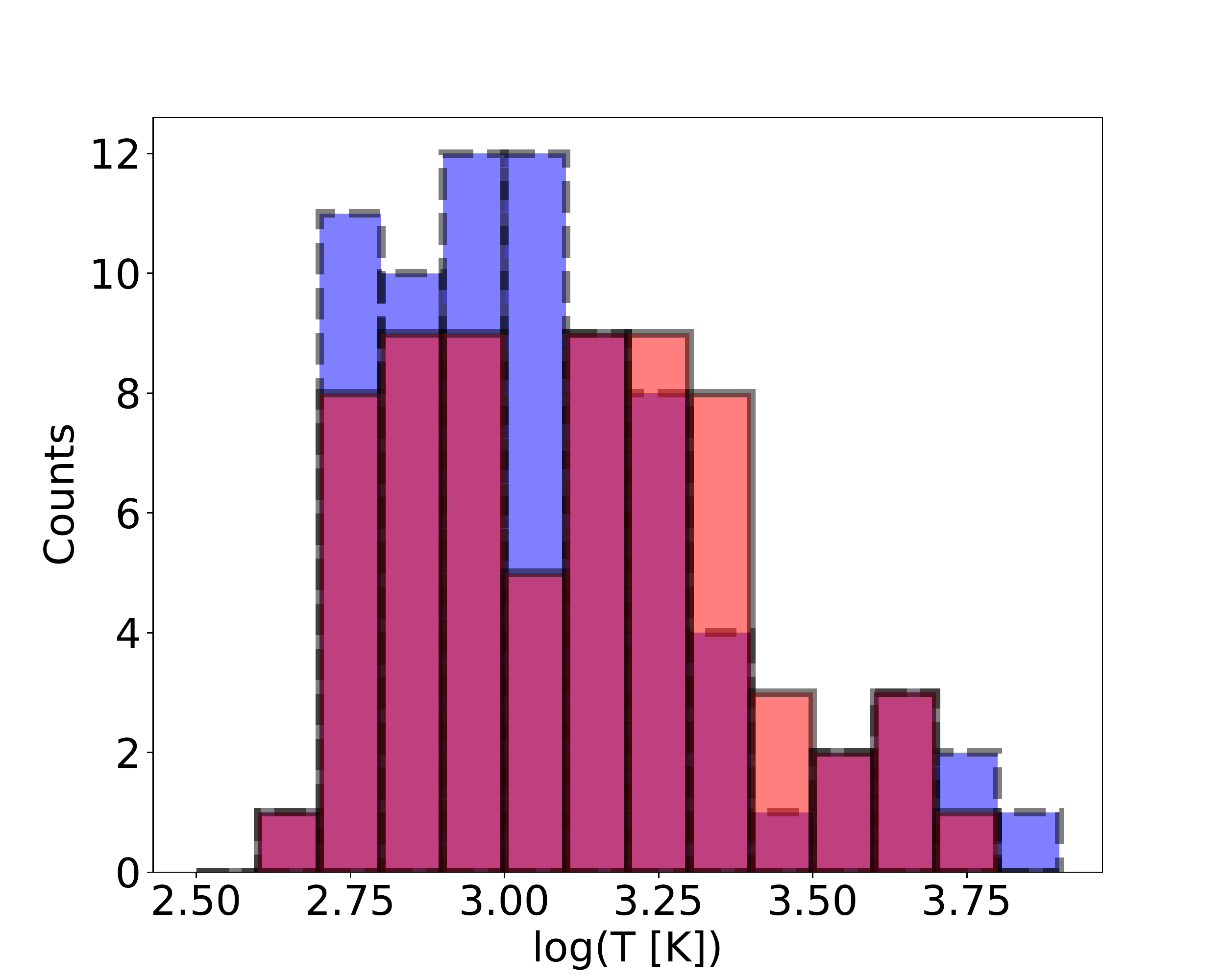}
    \includegraphics[width=.32\textwidth]{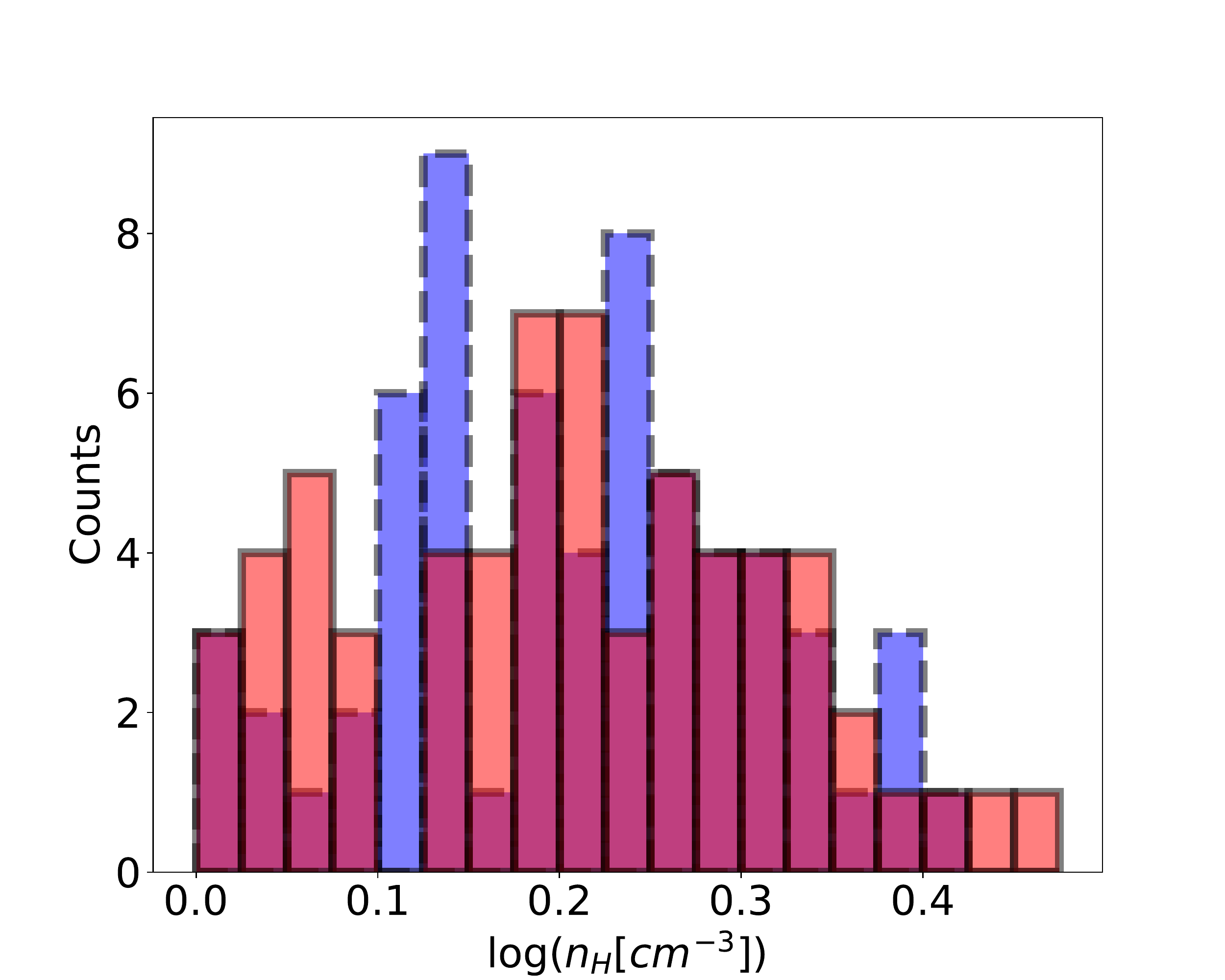}
    \caption{Histograms of SIGO properties in v2Cool (purple, dashed) and v2Uncool (pink, solid). Cooling does not dramatically affect SIGO properties. We consider the overdensity,  $\delta=(\rho_{\rm gas} -\bar{\rho})/\bar{\rho}$, top left panel, the prolateness,  ($\xi = R_{\rm max}/R_{\rm min}$), top right, the temperature bottom left, and lastly we show the number densities in the botom right.}
    \label{fig:SIGOhist}
\end{figure*}

The SIGOs overdensities are much larger than unity and thus, the SIGOs are in the process of collapsing \citep[as noted in][Figure 2]{popa}. We define overdensities as $\delta=(\rho_{\rm gas} -\bar{\rho})/\bar{\rho}$, where $\rho_{\rm gas}$ is the density of the gas cells composing the SIGO and  $\bar{\rho}$ is the average density of the Universe.  Interestingly we find that the large filaments around the SIGOs have overdensities that are much larger than unity, as depicted in Figure  \ref{fig:over}. 
While the typical size of a SIGO is about $10-50$~pc, it seems that in projection, already a larger regime (by a factor of a few) has an overdensity larger than unity, (see Figure \ref{fig:SIGOhist} for the 3-D overdensity of SIGOs). 
This is a complex neighboring environment, which includes both DM halos (with little to no gas component in them) and SIGOs in between (as shown in see Figure \ref{fig:SIGO_3d}). 

In Figure \ref{fig:over}, the black circles are projected DM halos in the vicinity of the SIGO. SIGOs tend to be found in the gas streams between DM halos. All SIGOs are collapsing at $z=20$, with the smallest SIGO having an overdensity of $\sim 400$ at $z=20$. We note that Figure \ref{fig:over} shows the results of the simulation with cooling. We find similar results without cooling.

\section{The relevance of cooling on SIGOs properties}
\label{sec:cooling}
\subsection{Physical properties}

It is well established that cooling affects the properties of DM/G objects \citep[e.g.,][]{katz+96, vogelsberger+12, Vogelsberger+13,hartwig+15} even in the presence of stream velocity \citep[e.g.,][]{schauer17, druschke+19, schauer+19}. We discuss these results in Appendix \ref{sec:appendixDMG}. In this section, we investigate the effect of cooling on SIGOs properties such as morphology, temperature, density, and environment.  Most of the previous investigations of SIGOs were conducted in simulations with no radiative cooling \citep{popa, chiou18}. In \citet{chiou+19}, for the first time, we studied the evolution of SIGOs while including  atomic cooling. Here we examine in details the effect that cooling has on SIGOs and compare to the no cooling realization.
Considering the temperature of SIGOs in the case of no cooling (v2Uncool), their temperatures range from $\sim400$~K to $\sim 6\times 10^3$~K (as shown in purple, dashed in the bottom left panel in Figure \ref{fig:SIGOhist}). Thus, since atomic cooling is rather inefficient for temperatures below $10^4$~K, we expect that atomic cooling will have little to no change on the SIGOs temperature, as shown in Figure \ref{fig:SIGOhist} bottom left panel. In fact, because of this inefficiency of atomic cooling, we hypothesize that other SIGO properties, such as overdensity, prolateness, and number density will be similar between the case of with and without cooling.
Histograms of these properties are shown in Figure \ref{fig:SIGOhist} on the top left, top right, and bottom right respectively. The KS tests we conducted between the two runs found that there was not enough evidence to conclude that the SIGO properties came from different distributions at 95\% confidence. The p-values, means, medians, and standard deviations can be found in the right column of Table \ref{table:properties}.

The distributions of overdensities (top left panel of Figure \ref{fig:SIGOhist}) are roughly normal in both runs, and all SIGOs have greater than $200$ overdensity.  The top right panel, depicts  the prolateness, defined by: $\xi = R_{\rm max}/R_{\rm min}$, where $R_{\rm max}$ ($R_{\rm min}$) is the largest (smallest) axis of the ellipsoid. While the prolateness is shifted to slightly higher values due to cooling, it is statistically indistinguishable from the non-cooling case. Similarly, the temperature distribution of SIGOs (bottom left panel) is shifted to slightly lower values, but again, as expected, is statistically indistinguishable from the non-cooling run. Furthermore, cooling does not affect the hydrogen number density (bottom right panel). Performing two-sample KS tests between v2Cool and v2Uncool all yield p-values $> .05$, implying that there is not enough information to reject the null hypothesis at $95\%$ confidence that these two simulations come from the same property distributions. 

\begin{figure*}
    \centering
\includegraphics[width=1\textwidth]{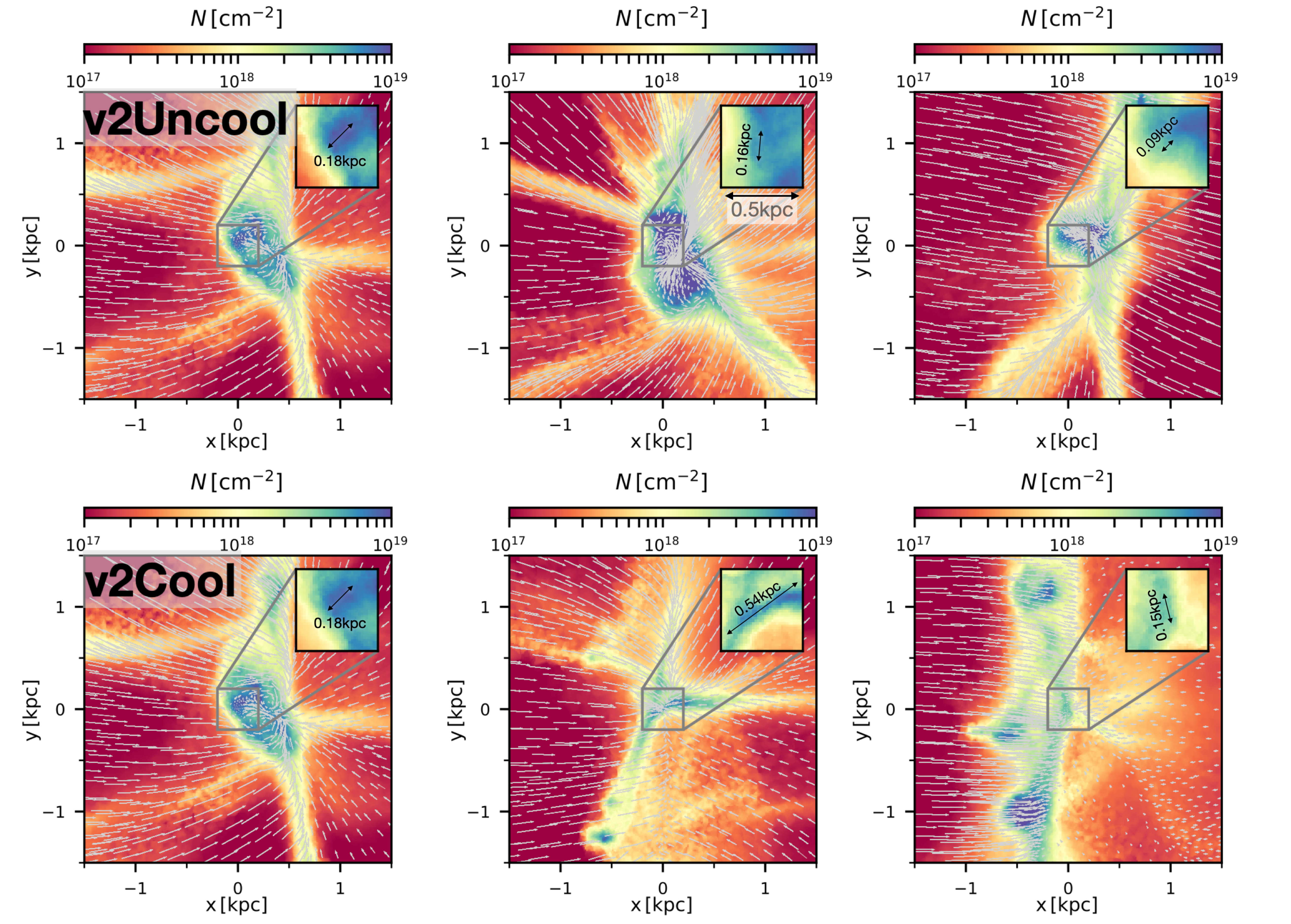}
    \caption{Column densities of a SIGO and its surrounding environment. A slice of the velocity field is also shown. Gas is accreting onto the SIGO. The top row contains SIGOs from the v2Uncool run and the bottom contains three from the v2Cool run. Also over-plotted in the zoom-in boxes are the representative large axis scales of the SIGOs, $L$. We caution that the scales on the SIGOs are here reported only for illustrative purposes.}
    \label{fig:SIGOdensity}
\end{figure*}

We note that in v2Cool we have removed one outlier SIGO that has an overdensity (density) of $1.3\times 10^6$ ($5.2\times 10^{-21}$~g~cm$^{-3}$), mass of $3\times 10^6$~M$_\odot$ and a temperature of 3100~${\rm K}$. It is 1.1 kpc away from the closest DM/G. There is no corresponding massive SIGO in v2Uncool \footnote{The most massive SIGO in the v2Uncool is $1.9\times 10^5$~M$_\odot$ \citep{chiou+19,popa}}. While massive, this is not unexpected. \citet{naoznarayan14} found an upper limit for a $2\sigma$ density fluctuation to be $\sim 10^6 \msun$. We further explore this object in Appendix \ref{sec:bigSIGO}. 

To illustrate further the consistency between the v2Cool and v2Uncool, we consider three representative examples of SIGOs in Figure \ref{fig:SIGOdensity}, with and without cooling (bottom and top panels, respectively). As can be seen in the Figure, the morphology, the environment, as well as the turbulent velocity field in the inner region of the SIGOs seem consistent between the two scenarios. This turbulent energy, driven by the supersonic stream velocity, must be overcome for star formation to occur. We note that the left panels correspond to the same SIGO between runs. In this Figure we zoom in to the innermost 0.5~kpc part of the SIGO as shown in the inset of each panel.
SIGOs in these simulations are highly nonlinear and dense objects, with a mean overdensity in v2Cool and v2Uncool of 657 and 690 and $n_H$ of 1.52 cm$^{-3}$ and 1.60 cm$^{-3}$ (see Table \ref{table:properties}). 

\begin{table*}
\begin{tabular}{|l|l|l|l|l|l|l|l|}
\hline
                  &      & v2Cool &          &      & v2Uncool &          & KS p-value \\ \hline
                  & mean & median & $\sigma$ & mean & median   & $\sigma$ &            \\ \hline
$\delta$          & 657  & 660    & 197      & 690  & 670      & 200      & 0.58       \\
$\xi$             & 8.33 & 7.23   & 4.30     & 9.47 & 7.81     & 8.79     & 0.50       \\ 
$T$ (K)           & 1429 & 1064   & 1127     & 1696 & 1421     & 1370     & 0.17       \\ 
$n_H$ (cm$^{-3}$) & 1.52 & 1.53   & 0.46     & 1.60 & 1.53     & 0.46     & 0.58      \\  
\hline \end{tabular} 
\caption{Table of summary statistics of SIGOs for v2Cool. We use a two sample Kolmogorov Smirnov (KS) test to compare the distributions of the SIGO parameters in v2Cool and v2Uncool.  Cooling does not significantly alter the distributions of SIGO properties. We note that in this statistics we dropped one major SIGO outlier, and we discuss its properties in appendix \ref{sec:bigSIGO}.}\label{table:properties}
\end{table*}

\subsection{The Effect of Cooling on Star Formation in SIGOs}

Star formation requires the gas within a DM/G or SIGO to be very cold and dense so that it can collapse gravitationally. In a previous paper \citep{chiou+19}, the cooling simulations that we ran did not explicitly follow star formation.
To study the star-forming properties, we used a simple stability argument similar to the Jeans criterion \citep{Jeans1902}, which describes the balance between gravity and thermal pressure. This criterion is often used in the literature for primordial star formation \citep[e.g.,][]{Bromm+99,Bromm+02}.
However, due to the supersonic nature of the stream velocity, turbulence must also be considered. Equating the Jeans length with the sonic length (i.e., the length scale at which gas transitions from supersonic to subsonic), \citet{Krumholz+05} defined a critical density for collapse, $\rho_{\rm crit}$. This critical density is given as 
\begin{equation}
\rho_{\rm crit} = \frac{\pi c_{\rm s}^2 \mathcal{M}^4}{GL^2},
\end{equation}
where $c_{\rm s}$ is the sound speed, $\mathcal{M}$ is the Mach number, $G$ is the gravitational constant, and $L$ is the scale at which turbulence is driven, and as demonstrated visually in Figure \ref{fig:SIGOdensity}, it coincides with the diameter of the SIGO along the longest axis. This critical density represents the balance between gravity and turbulent/thermal kinetic energy (in the absence of magnetic fields), above which the gas in the object will become unstable to collapse and the object will likely form stars. \citet{burkhart18a} related this critical density to the transition between a turbulent density distribution and one that is experiencing gravitational collapse and hence it is a natural consequence of the continuity of the density distribution function \citep[see also][]{girichidis+14, burkhart18b, guszejnov+18}. Here we use this critical density as an indicator of star formation.

\begin{figure}
    \centering
    \includegraphics[width=.5\textwidth]{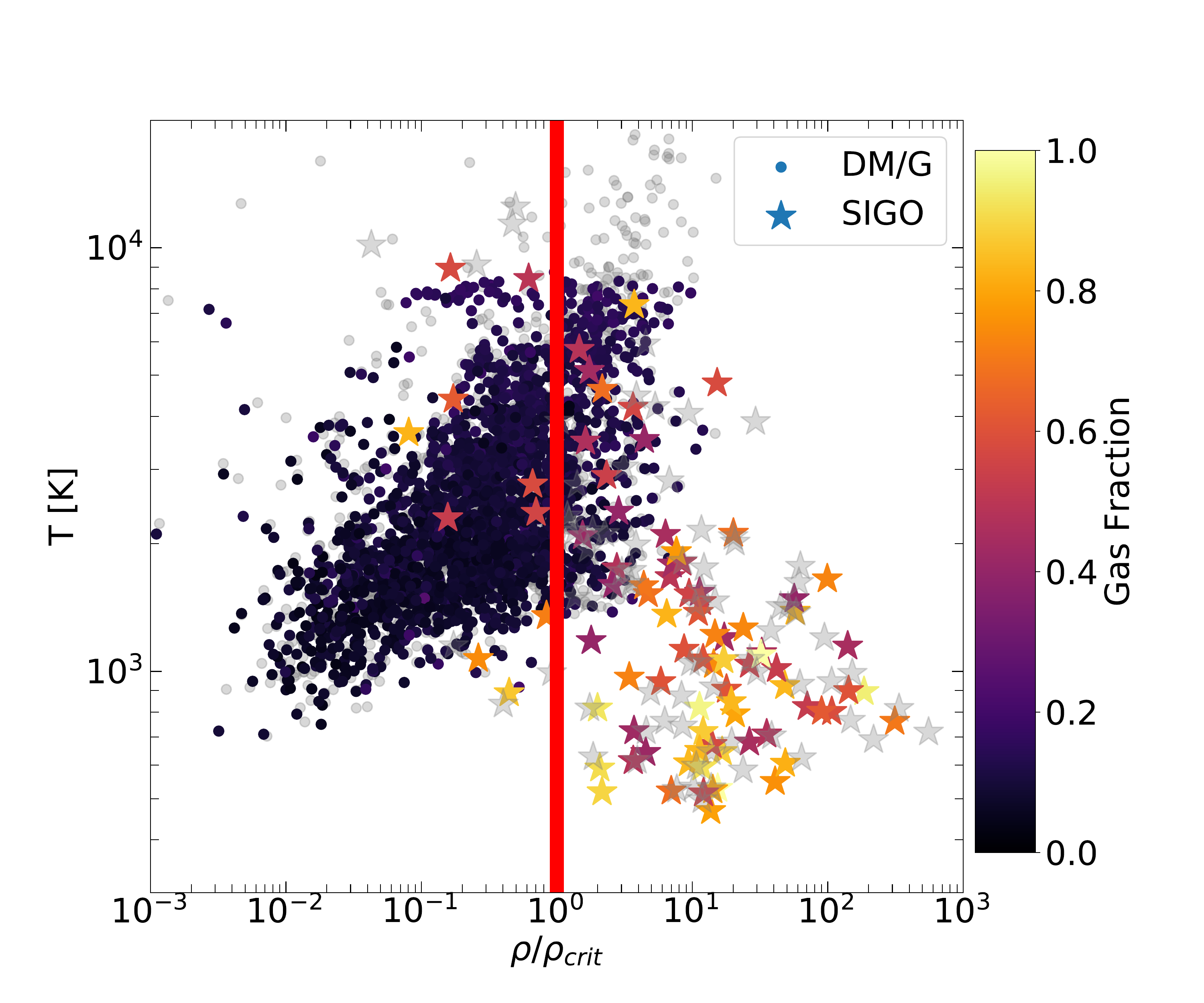}
    \caption{Temperature of objects as a function of as a function of $\rho/\rho_{\rm crit}$ at $z=20$ with stream velocity. All objects to the right of the red line are star forming (see \citet{chiou+19}. The gray points are from v2Uncool, whereas the colored points are from v2Cool. DM/Gs and SIGOs are colored by gas fraction. The temperatures decrease a bit due to cooling, however the overall shape of the distributions stay the same indicating that gravity is dominating.}  
    \label{fig:tempvsrho}
\end{figure}

\begin{figure}
    \centering
    \includegraphics[width=.5\textwidth]{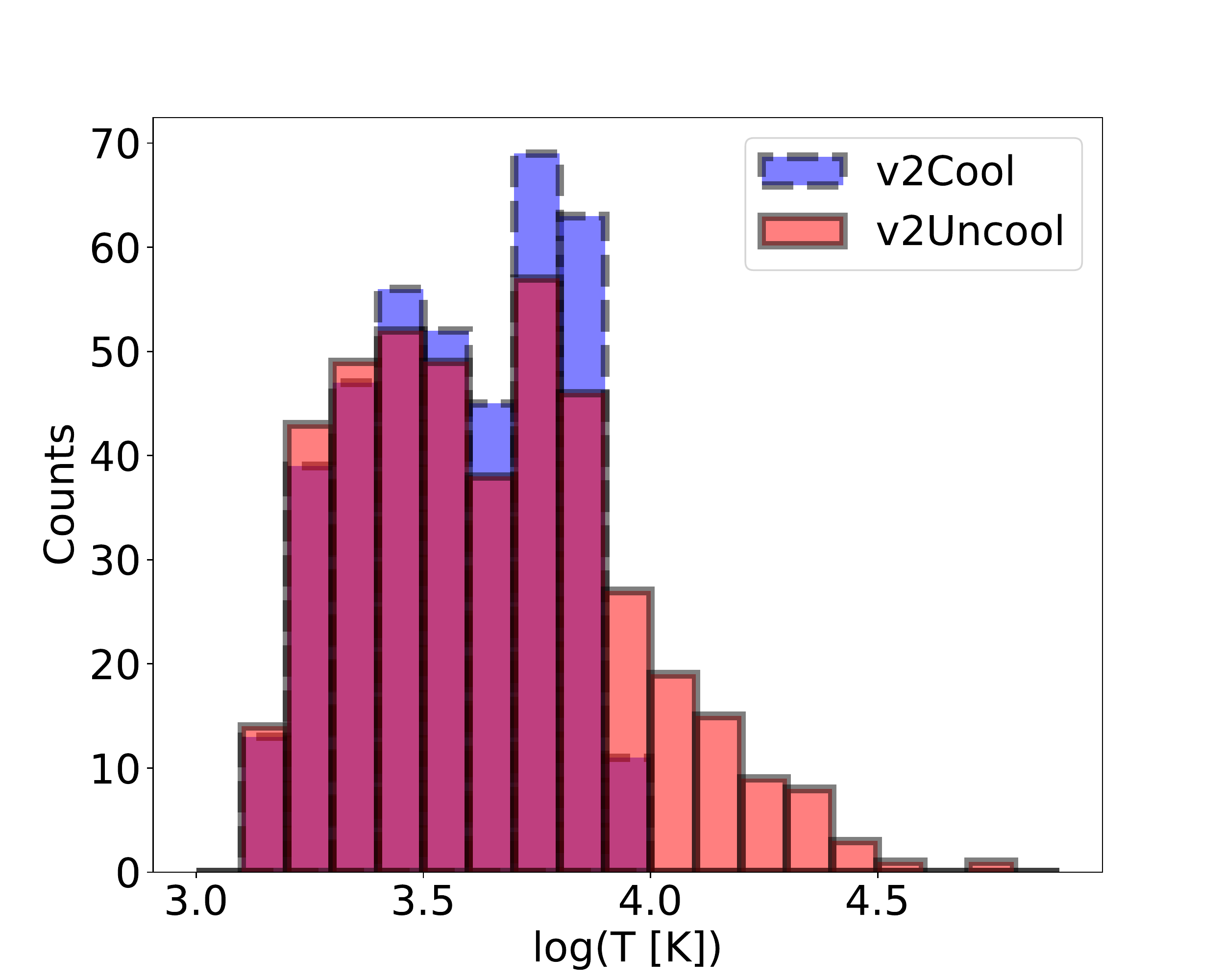}
    \caption{Histogram of temperatures of DM/Gs in v2Cool (purple, dashed) and v2Uncool (pink, solid). Atomic cooling is effective for temperatures greater than about $10^4$~K, hence there are no DM/G in v2Cool that have greater than this temperature.}
    \label{fig:DMGtemphist}
\end{figure}

In \citet{chiou+19} we have shown that SIGOs are ripe sites for start formation. Notably, $88\%$ of the SIGOs were star forming in the cooling simulations performed there (v2Cool). This is comparable to the v2Uncool run in which $91\%$ of SIGOs were star forming. Moreover, while classical small scale structures in the early Universe, i.e., DM/G, are less likely to form stars, SIGOs' critical density typically is much higher than unity \citep[as shown in][Figure 2]{chiou+19}.
In Figure \ref{fig:tempvsrho}, we show the cooling runs overlayed with the no cooling results in grey. The DM/G and SIGO points are colored by their gas fractions.  As expected, the cooling tends to lower the average temperatures to below $10^4~K$ of DM/G (Figure \ref{fig:DMGtemphist}), but the SIGOs' temperatures are not significantly affected. 

As can be seen in Figure \ref{fig:tempvsrho}, a few SIGOs in v2Uncool (grey stars at the left top corner - left to the red vertical line), have a temperature about $10^4$~K. SIGO counterparts in the v2Cool run indeed cooled via atomic cooling, reduced their temperatures and slightly increased their densities and hence are not present above $10^4$~K. We note that these SIGOs are not expected to form stars, because their density is smaller than $\rho_{\rm crit}$ unlike the majority of the SIGOs (located at the right of the red vertical line). Cooling does not significantly affect the scatter of SIGOs in the temperature - density plane.  Although the SIGOs in v2Uncool seem to achieve in general higher densities, it is not statistically significant (as shown in Table \ref{table:properties}). However, the scatter remains similar suggesting that gravity is the dominating mechanism for the onset of star formation in SIGOs (not so dissimilar to modern star formation \citep[e.g.,][]{burkhart18b}).

As highlighted here \citep[and in][]{chiou+19}, the turbulent medium that SIGOs live in show that star formation depends on the critical density threshold, $\rho_{\rm crit}$, \citep[as expected in the interstellar medium, e.g.,][]{burkhart18a}. Objects with densities above this critical density may undergo star formation. Given that $\rho_{\rm crit}$ depends on SIGOs morphology ($L$) and on velocities which did not significantly change, the value of $\rho_{\rm crit}$ remains similar with and without cooling (see Figure 5). Thus, it stands to reason that atomic cooling may not significantly affect the star-forming ability of SIGOs. Further inclusion of molecular hydrogen cooling should only increase the star formation potential of SIGOs.

\begin{figure}
    \centering
     \includegraphics[width=.5\textwidth]{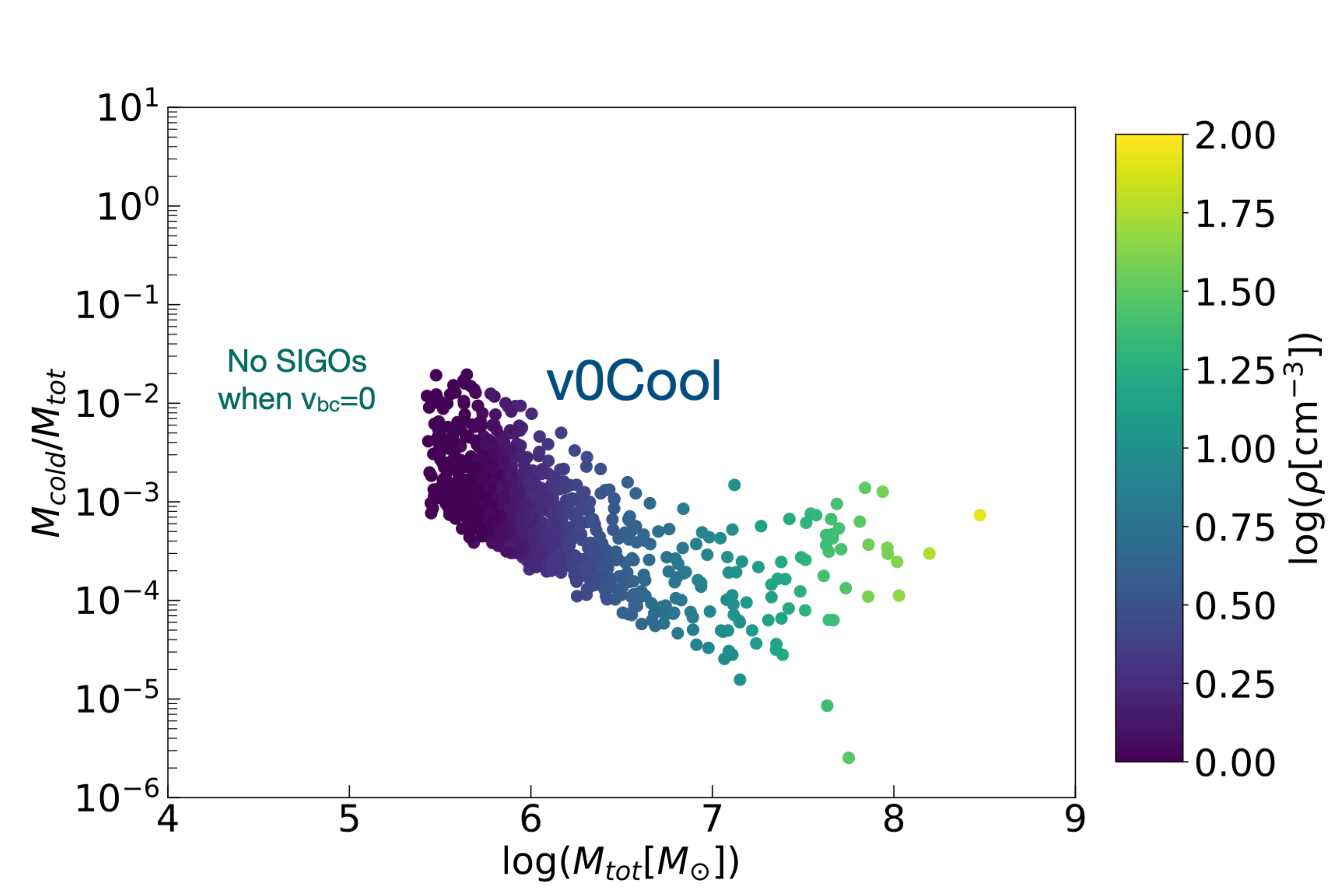}
    \includegraphics[width=.5\textwidth]{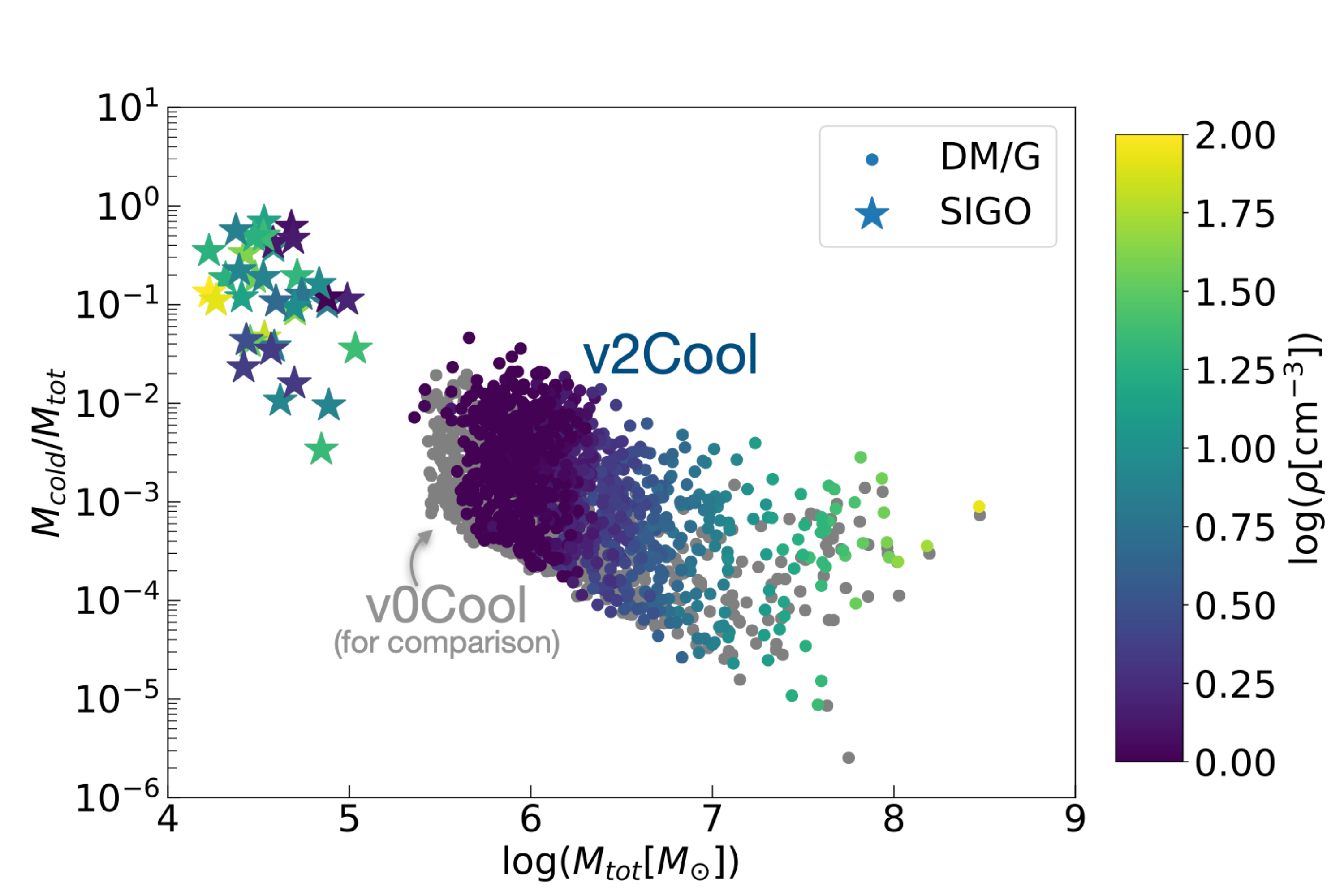}
       \caption{Ratio of cold gas mass to halo gas mass as a function of halo mass.
       We consider the $v_\bc=0$ simulation (v0Cool) in the top panel, where no SIGOs are present, and the $v_\bc=2\sigma_{\rm vbc}$ (v2Cool) in the bottom panel. For comparison we also plot the v0Cool simulation in grey in the bottom panel. The points are colored by the gas density of the object. 
       Here we define a cold gas cell to have a temperature of $<500$~K. The top panel corresponds to v2Cool and the bottom panel corresponds to v0Cool. For the v2Cool simulation, $29\%$ of DM/G and $43\%$ of SIGOs contain cold gas. In no stream velocity case, $24\%$ of DM/G contain cold gas. }
    \label{fig:cold}
\end{figure}

In the literature, the fraction of the cold gas in an object compared to the total gas fraction is often used as a tracer to star formation \citep[e.g.,][]{wang+17, davis+19, schauer+19}. Here, we define a cold gas cell to be one with a temperature $<500\,{\rm K}$ and define as a cold object a structure that contains at least one cold gas cell following \citet{schauer+19}.

In Figure \ref{fig:cold}, we plot the cold gas fraction, $f_{\rm cold}=M_{\rm cold} / (M_{\rm gas} + M_{\rm DM}) = M_{\rm cold} / M_{\rm tot}$, as a function of total mass of the object for v2Cool and v0Cool in the top and bottom panels. The points are colored by gas density. We find that $29\%$ of DM/G and $43\%$ of SIGOs contain cold gas in v2Cool. In v0Cool, only $24\%$ of DM/G contain cold gas (and, of course, no SIGOs are present). 

A clear feature for the zero stream velocity cooling run (v0Cool) is an increasing trend of cold mass fraction for masses $M_{\rm tot} \geq 10^7 \msun$. This kink corresponds to a virial temperature of about $10^4$~K. Above this temperature, atomic cooling  dominates \citep{barkanaloeb2001}. The stream velocity generally increases the scatter in the cold gas fraction in the DM/G. This is expected as the stream velocity acts as an additional pressure, enhancing collisions and diffusing the kink-feature at $10^7 \msun$. 

The SIGOs have overall higher cold gas fraction than the DM/G. Furthermore, as empathized in Figure \ref{fig:SIGOdensity},  SIGOs  generally have lower temperature and high gas densities compared to their DM/G counterparts.  Thus, it is not surprising that they present a high cold gas fraction (as depicted in Figure \ref{fig:cold}, top panel).

\section{Summary and conclusions}\label{sec:conclusions}
We have investigated the nature of Supersonically Induced Gas Objects (SIGOs) under the influence of atomic cooling in cosmological simulations. In previous simulations \citep[e.g.,][]{popa, chiou18} with only adiabatic cooling due to the expansion of the Universe, SIGOs were shown to be gas-rich ellipsoidal objects residing outside of dark matter halos. SIGOs at $z=20$ generally have overdensities well above 200 indicating collapse. 

We find that cooling does not dramatically alter the SIGO environment, though it does contribute to a more diffuse IGM, as shown in the density projections examples in Figure \ref{fig:SIGOdensity}. 
We find that there are no $\geq 3$ neighboring SIGOs within $\sim 2$~ckpc$^3$ box. However, rarely we find 2 neighboring SIGOs (see Figure \ref{fig:SIGO_3d}).  As was noted in \citet{chiou+19},  SIGOs live in highly turbulent environments  (e.g., Figure \ref{fig:SIGOdensity}) which assist with star formation.  

SIGOs form in the very early Universe, and are more abundant in higher density regimes or higher $\sigma$ density peaks \citep[][as we assumed here]{naoznarayan14}. At these early times the gas is pristine and therefore we consider the effects of atomic cooling only. 
However, atomic cooling is inefficient below $10^4$~K, thus, since SIGOs' temperatures due to adiabatic cooling is generally lower than $10^4$~K, we expect only a marginal effect of this cooling channel on SIGOs temperatures (see Figure \ref{fig:tempvsrho}). Similarly, the overdensities, prolateness, and hydrogen number densities are not substantially affected, as shown in Figure \ref{fig:SIGOhist}. The physics behind the SIGOs properties is, thus mostly determined by gravitational collapse and the stream velocity.

The combination of high densities and low temperatures yield promising nurturing ground for star formation for the SIGOs, as illustrated in Figure \ref{fig:tempvsrho} and in \citet{chiou+19}. The classical objects, (DM/G) on the other hand, not only have lower gas fractions due to the stream velocity effect, but also have too high a temperature ($T \sim 10^3-10^4~K$) to have efficient star formation.   
We speculate that molecular hydrogen cooling will further decrease the temperatures and clump SIGOs in a more pronounced way to produce more favorable sites for star formation which will be discussed in a future paper.

SIGOs have been purported as a novel formation channel for globular clusters \citep{naoznarayan14, chiou+19}. Here we showed that atomic cooling effects are negligible and the main contributor to SIGOs physical properties (such as temperature, density, etc.) is the underlining early Universe evolution \citep[as described in][]{naoznarayan14}. Thus,  semi-analytical studies can capture the essence of SIGOs as star-forming sites reliably. Nonetheless, additional physical processes in simulation, such as molecular cooling, star formation, and feedback, will allow for more accurate modeling of these possible progenitors. 
As the \textit{James Webb Space Telescope} comes online, future observations will further investigate the link between SIGOs and globular clusters.

\acknowledgments
We thank the referee for useful comments and questions. 
The authors thank Volker Springel for access to \textsc{arepo}. YSC, SN, BB, FM and MV acknowledges the partial support of NASA grant No. 80NSSC20K0500, and the XSEDE AST180056 allocation as well as the Simons Foundation Center for Computational Astrophysics for computational resources.
YSC thanks the partial support from UCLA dissertation year fellowship. SN  thanks Howard and Astrid Preston for their generous support.
FM is supported by the Program ``Rita Levi Montalcini'' of the Italian MIUR.
BB is grateful for the support from the Simons Foundation Center for Computational Astrophysics.

\vspace{5mm}
\appendix

\section{The biggest SIGO}\label{sec:bigSIGO}
\begin{figure*}
    \centering
    \includegraphics{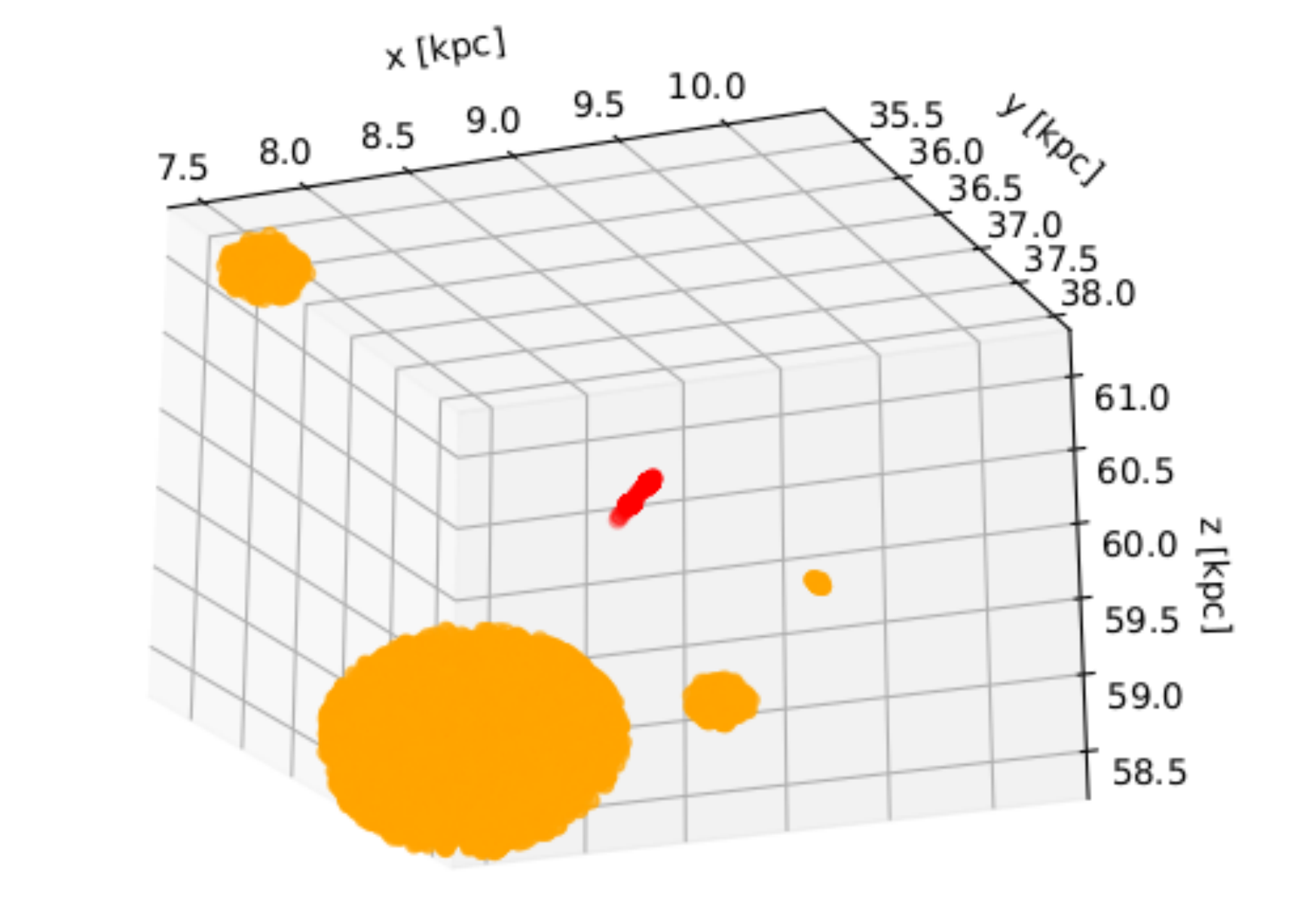}
    \caption{The largest SIGO (red) in v2Cool with mass $M_{\rm tot} \sim 3\times 10^6$~M$_\odot$ Surrounding DM/G are in orange. The largest DM/G shown here is the second most massive in the simulation. The SIGO is 1.1~kpc from the closest DM halo.}
    \label{fig:bigSIGO}
\end{figure*}

In this section we discuss the properties of the outlier SIGO in v2Cool shown in Figure \ref{fig:bigSIGO}. The SIGO (in red) is 1.1~kpc away from the closest DM/G, clearly out of the virial radius. Surrounding DM/Gs are shown in orange. The largest DM/G displayed here is the second largest in the simulation. This SIGO was the most dense ($5.2\times 10^{-21}$~g~cm$^{-3}$) and massive ($3\times 10^6$~M$_\odot$) SIGO found in v2Cool, an order of magnitude more massive than the most massive found in V2Uncool (which was $1.9\times10^5\msun$). It was also an order of magnitude more massive than the next most massive SIGO in v2Cool. The temperature was 3100~${\rm K}$. 

Using linear theory, \citet{naoznarayan14} showed that a $2\sigma$ overdense patch could form SIGOs as big as $\sim 10^6 \msun$.  Although there is no corresponding object in v2Uncool, we conjecture that this object is a result of Poisson statistics (small number statistics, of such a large object in a small box) and not a  result of cooling.

\section{DM/G} \label{sec:appendixDMG}
In this section we briefly reproduce several aspects of the effects of cooling on classical DM halos such as density projections and temperature distributions. 

\begin{figure*}
    \centering
    \includegraphics[width=1\textwidth]{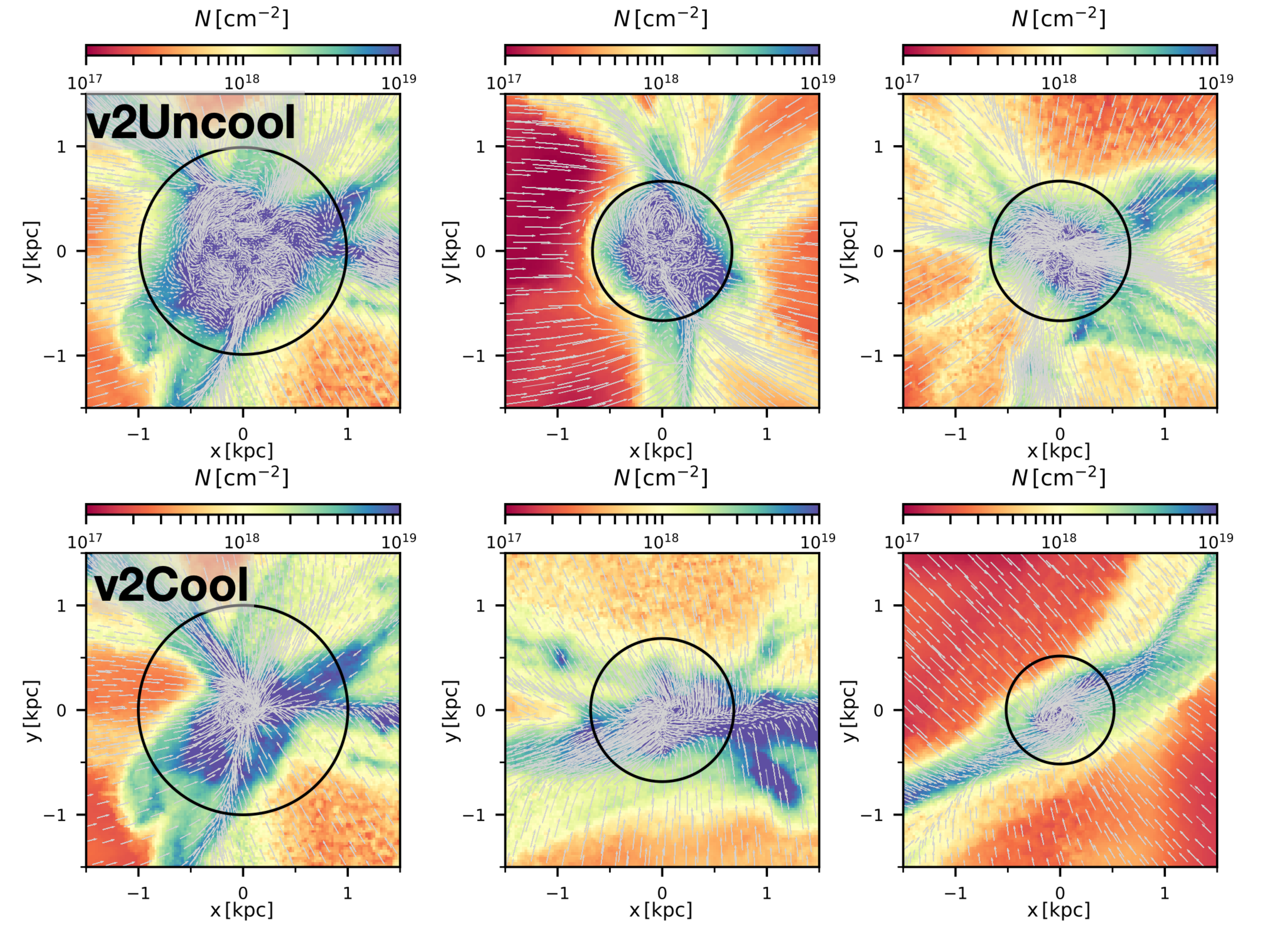}
    \caption{Column densities of a DM/G and its surrounding environment. A slice of the velocity field is also shown. Gas is accreting onto the DM/G. The top row contains DM/Gs from the v2Uncool run and the bottom contains three from the v2Cool run. The black circle indicates $R_{200}$ of the DM/G. }
    \label{fig:DMGdensity}
\end{figure*}

Several gas density projections of DM/Gs are shown in Figure \ref{fig:DMGdensity}. The top panels are from the v2Uncool run and the bottom are from v2Cool run. The circles represent the virial radius of the DM/G. The white arrows are velocity maps showing accretion onto the DM/G. The left DM/G is the most massive DM/G in each simulation and has similar large scale features. The cooling serves to compactify the DM/G as expected. The outer environments of the DM/G are less dense overall due to the gas condensing on the halo. The velocity field within the DM/Gs is more turbulent in the v2Uncool run since in general the DM/Gs in v2Cool have lower average temperature. 

\bibliography{myBib}{}
\bibliographystyle{aasjournal}

\end{document}